\documentclass[12pt]{article}

\def\be{\nopagebreak[3]\begin{equation}}

\def\ee{\end{equation}}

\def\ba{\nopagebreak[3]\begin{eqnarray}}

\def\ea{\end{eqnarray}}

	\def\ni{\noindent}

\begin{document}

	\begin{center}

	\centerline{}

	\vspace{.5cm}

	{\large\bf Hamiltonian and Volume Operators \\}

	\vspace{0.7cm}

	{\large\em M. Pierri\footnote{pierri@fma.if.usp.br}\\}

	\vspace{0.5cm}

	{ Instituto de F\'{\i}sica, Universidade de S\~ao Paulo  \\

	Caixa Postal 66318, 05315-970, S\~ao Paulo, SP, Brasil\\ }

	\vspace{0.5cm}

	\end{center}

\begin{abstract}
2+1 gravity coupled to a massless scalar field has an initial singularity when the spatial slices are compact. The quantized model is used here to investigate several issues of quantum gravity. The spectrum of the volume operator is stu\-died at the initial singularity. The energy spectrum  is obtained.  Dynamics of the  universe is also investigated. 

\vspace{0.3cm}

\end{abstract}

\section{Introduction}\label{i}

When  a consistent theory of quantum gravity is  rigorously formulated we expect it  to provide answers for all  existing open questions in the field. While this is not achieved, a natural route is to investigate these issues on symmetry reduced models of general relativity. They have less degrees of freedom but still keep some of the features of the full theory.  The quantum theory of 2+1 gravity coupled to an axi-symmetric massless scalar field is such a model. It  provides, also, an arena to investigate the interaction of gravity with matter fields. When the spatial slices are non-compact, this model was rigorously quantized and carefully analyzed on \cite{am96}. In \cite{aa96}, Ashtekar investigates quantum fluctuations of the geometry in view of  semi-classical issues. 

The compact case was consistently quantized in \cite{m01}. Due to  compactness, the spacetime has an initial singularity that   persists in the quantum theory \cite{m01,v87}. The model is now described by operators acting on a Hilbert space. There are several questions to be asked, as for instance,  is the geometry discrete close to the singu\-la\-ri\-ty? Is the singularity avoided in any sense? The Hamiltonian operator comes from deparametrization and has no physical meaning a priori. Indeed the vacuum is not an eigenvector with zero eigenvalue. Is it possible to obtain a new representation such that the new vacuum corresponds to a zero eigenvalue for the Hamiltonian operator?  It is also interesting to investigate the spectrum of the Hamiltonian operator and consequently the dynamical evolution of the system. The purpose of this note is to investigate these questions.

We start with a review of the quantized model. In section \ref{ns} a new representation is introduced that will greatly simplify the analysis of the spectrum of the Hamiltonian operator. The diagonalization of  the Hamiltonian operator is obtained in the next section.  Dynamics of the early space-time will be investigated. The solutions for the  Schr\"odinger equation  are explicitly obtained in the limit of  small and large times. The volume operator is analyzed in section \ref{vo}. Since this model is completely equivalent to the Gowdy ${\rm T}^3$ model \cite{g71} in 4-dimensional vacuum general relativity, we will make comparisons whenever it is appropriate. Finally, in section \ref{s}, we summarize the main results.

\section{Quantum Theory}\label{qm}

In this section we will summarize the quantum theory of 2+1 gravity on $M\equiv {\rm R}\times {\rm T}^2$ coupled to an axi-symmetric massless scalar field $\psi $. For more details we refer the reader to ref.\cite{m01}.

As stated in \cite{m01} the problem decouples, i.e.,  we first quantize  the scalar field on a flat background and then obtain the metric operator as a derived quantity. This model has a global constraint. Therefore the physical Hilbert space corresponds to the kernel of the global constraint.  Due to compactness of the spatial slices $\Sigma = {\rm T}^2$ there is a global degree of freedom $(q,p)$ whose quantization is trivial. We denote by $\stackrel{\circ}{\cal H}$ the Hilbert space were $\hat q$ and $\hat p$ are promoted to well-defined operators.

Let us consider a fiducial symmetric Fock Space ${\bar{\cal F}}$ where the following decomposition into creation and annihilation operators is naturally well defined

\be
{\hat \psi}(\theta,T)= \sum_{m=-\infty}^\infty \left( f_m(\theta,T){\hat A}_m +  \right.
\left.f_m^*(\theta,T){\hat A}^{\dagger}_m  \right)
\label{1}
\ee

\ni
where $\theta$ is an angular coordinate  with range $0\leq\theta<2\pi$ and

\ba 
f_0 (\theta,T) &=& \frac{1}{2}\left( \ln T -i\right)\nonumber \\
f_m (\theta,T) &=& \frac{1}{2}H_0^{(1)} (|m|T)e^{ im\theta} \nonumber \\ 
&=& \frac{1}{2}\left( J_0(|m|T) + i N_0 (|m|T)\right) e^{ im\theta}
\quad {\hbox{for}}\quad m\neq 0
\label{2}
\ea

\noindent
here $H_0^{(1)}=J_0 + iN_0$ is the 0th-order Hankel function of the 1st kind 
and $J_0$ and $N_0$ are the 0th-order Bessel function of the first and 
second kind respectively. (* denotes complex conjugation).  

The space ${\cal F}_{\sl phys}$ of physical states   is the subspace 
of the fiducial Hilbert space ${\cal F}=\stackrel{\circ}{\cal H}\otimes{\bar{\cal F}}$ defined by

\be
:{\hat P}_\theta:|\Psi>_{\sl phys}=0. 
\label{3}
\ee

\ni
where 

\be
P_\theta:= \int_0^{2\pi} 
      d \theta\, 2T:{\hat{\dot\psi}}(\theta) {\hat\psi'}(\theta):=2{\sum_{m=-\infty}^\infty}
   {\hat A}^{\dagger}_m {\hat A}_m.         
\label{4}
\ee

\ni
Thus, the physical states are  states such that   the
total angular momentum in $\theta$-direction vanishes.  

A generic physical state with N particles is given by

\be
|{}^N\Psi>_{\sl phys} = |\phi> \otimes\left[\prod_{i=1}^N |m_i>\right] \quad{\hbox{such %%@
that}}\quad
      \sum_{i=1}^{N} m_i =0.
\label{5}
\ee

\ni
where  $|\phi>$ is a state that belongs to $\stackrel{\circ}{\cal H}$ and $|m_i>$ represents a 
one-particle state with angular momentum $m_i$.

Dynamics is generated by the following Hamiltonian operator

\be
{\hat H}(T) = \int_0^{2\pi}d\theta 
      :{\hat p}T\left( ({\hat {\dot\psi}})^2 +   ({\hat \psi}')^2 \right):\, ,
\label{6}
\ee

\ni
or expressed in terms of creation and annihilation operators,

\ba 
{\hat H} &=& \frac{\pi}{2T}:\hat{p}({\hat A}_0 + {\hat A}^\dagger_0)^2: \nonumber\\
&+& \frac{T}{4} \sum_{m=-\infty}^{\infty} m^2  \hat{p}          
\left[ 2 {\hat A}^{\dagger}_m {\hat A}_m (H_1^{(1)}H_1^{(2)}+ %%@
H_0^{(1)}H_0^{(2)})\right.\nonumber\\
&+& {\hat A}_m{\hat A}_{-m} \left( (H_0^{(1)})^2+  (H_1^{(1)})^2\right)
+ \left.{\hat A}^{\dagger}_m {\hat A}^{\dagger}_{-m} \left( (H_0^{(2)})^2+ %%@
(H_1^{(2)})^2\right)\right],
\label{7}
\ea

\ni
where $H_0^{(2)}(|m|T)=[H_0^{(1)}(|m|T)]^*$ and $H_1^{(1,2)}(|m|T)=-\frac{1}{|m|}{\dot H}_0^{(1,2)}(|m|T)$.

The Hamiltonian operator commutes with the global constraint ${\hat P}_\theta$. Therefore  it is  also a well-defined operator on the physical Hilbert space ${\cal F}_{\sl phys}$. Note that the Hamiltonian is time-dependent. 

The spacetime quantum geometry is recovered with the following regularized me\-tric operator

\be
{\hat g}_{ab}(f_\theta) =  e^{\hat{q}}e^{ \int_0^{2\pi} 
f_\theta(\theta_1)2T :{\dot{\hat \psi}}{\hat \psi'}:d\theta_1}
\left(-\nabla_a T\,\nabla_b T\, + \nabla_a \theta\nabla_b \theta \right) 
+ {\hat{p}^2}T^2 \nabla_a\sigma\, \nabla_b\sigma\, ,
\label{8}
\ee

\ni
where the regulator $f_\theta(\theta_1)$  equals $1$ for 
$\theta_1\leq \theta- \epsilon$, then it smoothly decreases to zero 
and equals zero for $\theta_1 \geq \theta-\epsilon$. Also,  
$f_{2\pi}(\theta)=f_0(\theta)$. 

Although we have obtained a well-defined operator on $\bar{\cal F}$, 
it is not an operator on the physical space because it 
does not commute with the global constraint ${\hat P}_\theta$.  
However,  it can be projected to the 
physical space. We will denote the projection operator by ${\cal P}$. Therefore the physical regulated metric operator
in ${\cal F}_{\sl phys}$ is given by: 

\be
[{\hat g}_{ab}(f_\theta)]_{\sl phys}= {\cal P} \,{\hat g}_{ab}(f_\theta)\,{\cal P}.
\label{9}
\ee

There are states, $|\phi_c>$, in  $\bar{\cal F}$ yielding semi-classical geometry. They are the usual coherent states for the field operators. The expectation value of an operator in such a state gives the classical expression. The coherent states can be projected to the physical Hilbert space. This construction can be found in \cite{m01}.

Since classically this spacetime has an initial singularity, the coherent states provide an example to show that it persists in the quantum theory. We will discuss in more details the singularity later on in this paper. 

One can show that the vacuum expectation value of the metric operator is a flat metric that has an initial singularity. The expectation value of curvature invariants would be vanishing on vacuum. Therefore the spatial volume operator is more appropriate to investigate the singularity. For $T\rightarrow 0$ it is given by,

\be
{\hat V}=\int_o^{2\pi}\int_o^{2\pi}\sqrt{h}d\theta d\sigma=2\pi 
{\hat p}Te^{{\hat q}/2}\sum_{j=0}^{+\infty} \frac{1}{j!}(\frac{ln T}{2})^j\int_0^{2\pi} d\theta \left(
\frac{1}{2}({\hat A}_0 +{\hat A}_0^\dagger)+{\hat f}(\theta))\right)^{2j} 
\label{9a}
\ee

\noindent
where $h$ is the determinant of the spatial metric and ${\hat f}(\theta)=\frac{i}{\pi}\sum_{m\neq 0}({\hat A}_m e^{i m \theta} - {\hat A}_m^\dagger e^{-i m \theta})$

\section{New Set of Creation and Annihilation Operators}\label{ns}

We now consider a new set of creation and annihilation operators that will greatly simplify the dynamics of the system

\ba
{\hat a}_0 & = & -i}{\sqrt{\frac{\hat p}{2}}(\sqrt3{\hat A}_0 + {\hat A}^\dagger_0)\nonumber\\
{\hat a}_n &=& \sqrt{\frac{{\hat p}}{\sqrt\pi}}\frac{\alpha_n{\hat A}_n+ (\beta_n-\frac{|n|}{\sqrt\pi}){\hat A}^{\dagger}_{-n}}{\sqrt{|n|(\beta_n-\frac{|n|}{\sqrt\pi})}}\quad{\hbox{for}}\quad n\neq 0 
\label{10}
\ea

\ni
where $\alpha_n(T)=\frac{T}{4}n^2[(H_0^{(1)})^2+(H_1^{(1)})^2]$ and $\beta_n(T)=\frac{T}{4}n^2[H_1^{(1)}H_1^{(2)}+H_0^{(1)}H_0^{(2)}]$.
The new set satisfies the commutation relations $[{\hat a}_n,{\hat a}^\dagger_{n'}]=\delta_{nn'}$.

The Hamiltonian operator (\ref{7}) expressed in terms of the new set yields

\be
:{\hat H}:= \frac{1}{T}{\hat P}_0^2+\sum_{n=-\infty}^{\infty}|n|({\hat a}^\dagger_n{\hat a}_n)
\label{11}
\ee

\ni
where  ${\hat P}_0= \frac{(\sqrt3-1)\sqrt{\pi}}{2}[i({\hat a}^\dagger_0-{\hat a}_0)]$. Thus, the zero mode corresponds to a  particle and the other n-modes are simple harmonic oscillators. Note that the new vacuum defined by ${\hat P}_0\otimes{\hat a}_n\otimes {\hat q}|0>=0$, 
$\forall\; n\neq0$  is an eigenvector of the Hamiltonian with zero eigenvalue. As can be easily verified from the Hamiltonian expressed in (\ref{7}), the  `old' vacuum did not satisfy this requirement.  
The spectrum of the Hamiltonian operator for each instant of time is straightforward.

The global constraint in terms of the new operators still has the same simple form. The physical states are such that

\be
:{\hat P}_\theta: |\psi>_{\sl phys}=\sum_{n=-\infty}^{+\infty}n{\hat a}^\dagger_n{\hat a}_n|\psi>_{\sl phys}=0
\label{12}
\ee

\section{The Schr\"{o}dinger Equation}\label{se}

Let us investigate the Schr\"{o}dinger equation associated with (\ref{11}) in the limit $T\rightarrow 0$.

\be
i \frac{\partial}{\partial T}|\Psi>_{\sl phys}(T) = {\hat H}(T) |\Psi>_{\sl phys}(T).
\label{13}
\ee

\ni

The Hamiltonian close to the singularity is of the form: 

\be
{\hat H}(T)=\frac{1}{T}
\left({\hat H}^{(0)}+ \sum_{n=-\infty}^{+\infty}{\hat H}^{(n)}\right)=\frac{1}{T}
\left({\hat P}_0^2+\sum_{n=-\infty}^{\infty}|n|({\hat{\stackrel{\circ}{a}}}^\dagger_n
{\hat{\stackrel{\circ}{a}}}_n)\right). 
\label{13a}
\ee

\ni
where ${\hat{\stackrel{\circ}{a}}}^\dagger_n$ and ${\hat{\stackrel{\circ}{a}}}_n$ are time independent operators. The time dependence left is $1/T$ only.
Therefore the solution for the Schr\"odinger equation is 

\be
|\psi(q_0,t)> = \left( \int g(k,t_0) e^{i (kq_0- \omega t)}dk\right)\otimes \left(\prod_{n=-\infty}^{+\infty}\sum_{N_n}e^{-iE_{N_n}t} c_{N_n}(t_0)|\phi_{N_n}>\right)
\label{14}
\ee

\noindent
where  $t=ln T$, $\omega=k^2$, $E_{N_n}=|n|N_n$ and ${\hat H}^{(n)}|\phi_{N_n}>=E_{N_n}|\phi_{N_n}>$. $g(k,T_0)$ and $c_{N_n}(T_0)$ are determined from the initial data. The solution of the Schr\"odinger equation is a free particle for the zero mode and harmonic oscillators for all other modes. 

Moreover the metric operator (\ref{8}) becomes

\be
{\hat g}_{ab} = T^{{\hat k}(\theta)}e^{\hat q}(-\nabla_a T \nabla_b T +\nabla_a  \theta \nabla_b \theta)
+  p^2T^2 \nabla_a \sigma \nabla_b \sigma.
\label{16}
\ee

\noindent
where  ${\hat k}(\theta)=\frac{1}{2}\frac{{\hat P}_0}{\sqrt{2\pi{\hat p}}}+{\hat f}(\theta)$.
Therefore, after redefining the time coordinate,   the space-time behaves as some Kasner solution for each value of $\theta$. This result is also obtained from the 4-dimensional perspective \cite{b74}.

For large $T$ the time evolution of the zero mode becomes negligible (\ref{11}). Therefore it behaves as a frozen degree of freedom. The non trivial part of the Hamiltonian of the system consists of harmonic oscillators for all modes except the zero mode and is time independent, i.e., 

\be
{\hat H}=\sum_{n=-\infty}^{+\infty}{\hat H}^{(n)}=
\sum_{n=-\infty}^{\infty}|n|({\hat{\stackrel{\infty}{a}}}^\dagger_n
{\hat{\stackrel{\infty}{a}}}_n).
\label{16a}
\ee 

\ni
where ${\hat{\stackrel{\infty}{a}}}^\dagger_n$ and ${\hat{\stackrel{\infty}{a}}}_n$ are time independent operators. The Schr\"odinger equation can be solved  in this limit and yields 
\be
|\psi(T)>=\prod_{n=-\infty}^{+\infty}\sum_{N_n}e^{-iE_{N_n}T}c_{N_n}|\phi_{N_n}>
\label{17}
\ee

This state can be projected to the physical subspace. Moreover the usual coherent states for the harmonic oscillators can be constructed and physical coherent states as well. This was performed  in detail in Ref. \cite{m01}. The same construction can be applied here since the global constraint still has the same structure on the new set of creation and annihilation operators as before.

Note that the states do not have a singularity on $T=0$ anymore, nor does the Hamiltonian. Therefore the quantum system for large $T$ does not show the presence of the singularity at $T=0$.

 A remark is in order here. The simple form of the Hamiltonian achieved here is due to our choice of parametrization for the system.  In \cite{m73}, the limit of large $T$ yields a Hamiltonian that is  a sum of time dependent harmonic oscillators.

\section{Volume Operator}\label{vo}

In this section we will obtain the spectrum of the volume operator in the limit $T\rightarrow 0$. 

As it was seen in the previous section, at both limits the n-modes  ($n\neq 0$) behave as harmonic oscillators. Whereas the zero mode is relevant close to the singularity and is `frozen' for large $T$. Therefore without loss of generality we will assume a coherent state for all modes except the zero. Furthermore we will compute the expectation value of the volume operator in such a state. 

In terms of the new set of creation and annihilation operators the eigenvalue equation for the volume operator (\ref{9a}) is explicitly

\be
{{\hat V}_c}^{(0)}|\psi(T)>=2\pi p Te^{q/2}\sum_{j=0}^{+\infty} \frac{1}{j!}(\frac{ln T}{2})^j\int_0^{2\pi} d\theta \left(
\frac{{\hat P}_0}{\sqrt{2\pi p}} +f_c(\theta)\right)^{2j}|\psi(T)> 
\label{18}
\ee

\ni
where ${{\hat V}_c}^{(0)}=<\psi_c^{(n)}|:{\hat V}:|\psi_c^{(n)}>$ is the expectation value of the volume operator in a  state that is the tensor product of the coherent state for all modes $n\neq 0$ and a state where ${\hat p}$ and ${\hat q}$ are peaked on their classical value. The spectrum of ${{\hat V}_c}^{(0)}$ is continuous. As can be seen  in the momentum representation 

\be
{{\hat V}_c}^{(0)}\psi(p_0, T)=2\pi  pTe^{q/2}\sum_{j=0}^{+\infty} \frac{1}{j!}(\frac{ln T}{2})^j\int_0^{2\pi} d\theta \left(
\frac{P_0}{\sqrt{2\pi p}}+{f_c}(\theta)\right)^{2j}\psi(p_0,T)
\label{19}
\ee

\ni
Generically, the singularity at $T=0$ persists. The volume of the universe vanishes at this instant of time. 

Let us consider the leading order in $T$. It is interesting to note that there is a sector of the Hilbert space such that the norm of the volume operator  is finite and different from zero at $T=0$, specifically

\be
\int_{-\infty}^{+\infty} dq_0|{{\hat V}_c}^{(0)}\psi(q_0,T)|^2=(2\pi p)^2e^q
\label{20}
\ee

\ni
The frequency of states belonging to this sector satisfy the relation $\omega+\sum_n\sum_{N_n}E_{N_n}=-i$. Furthermore, if we assume vacuum for all modes $n\neq0$, solutions of the free particle in this sector of the Hilbert space will yield finite norm considering all terms in the expression (\ref{19}).

\section{Summary}\label{s}
The Hamiltonian operator is cast in a simple form when expressed in terms of the new set of creation and annihilation operators. The energy spectrum, for each instant of time, is a sum of a `free particle' and harmonic oscillators. The spectrum closer to the singularity is the sum of the energy of a `free' particle for the zero mode and time independent harmonic oscillators for all other modes with an overall factor of $T^{-1}$. Far away (timewise) from the initial singularity, the zero mode is frozen in the sense that its evolution is negligible in comparison with the other modes. The energy spectrum, then,  is a sum over all modes  of the  energy of time independent harmonic oscillators. Solutions for the Schr\"odinger equation is obtained in the small and large time regimes. 
   
The spectrum of the volume operator is continuous close to the singularity. Gene\-ri\-cally the singularity persists.  There is a sector of the Hilbert space that yields a non-vanishing norm for the volume operator. The energy of states belonging to this sector satisfies a complex constraint. 

\section{Acknowledgments}
I would like to thank A. Ashtekar, A. Gadidov and P. Teotonio-Sobrinho for discussions.   This work was supported by FAPESP.

\end{document}